\documentclass[12pt]{iopart}
\usepackage{iopams}
\usepackage{theorem}
\usepackage{graphicx}

\newcommand\cG{{\mathcal G}}

\newcommand\cM{{\mathcal M}} 
\newcommand\cN{{\mathcal N}}

\newcommand\field{\mathbb}
\newcommand\R{\field{R}}

\newcommand\C{\field{C}}
\newcommand\Z{\field{Z}}
\newcommand\N{\field{N}}

\newcommand\Dt{\frac{\rmd\phantom{t} }{\rmd t}} 
\newtheorem{theorem}{Theorem}
\newtheorem{lemma}{Lemma}

\newtheorem{remark}{Remark}
\newenvironment{proof}{\textit{Proof.}}{\hfill$\Box$}

\begin{document}
\title[Non-integrability of the  generalised spring-pendulum problem]{Non-integrability of the  generalised spring-pendulum problem}
\author{ Andrzej J.~Maciejewski\dag\,  Maria Przybylska\ddag\S\ and Jacques-Arthur Weil\ddag\P}
\address{\dag\ Institute of Astronomy,
  University of Zielona G\'ora,
  Podg\'orna 50, 65--246 Zielona G\'ora, Poland}
\address{\ddag\ INRIA -- Projet {\sc  Caf\'e},
2004, Route des Lucioles, B.P. 93,
06902 Sophia Antipolis Cedex, France}
\address{\S\  Toru\'n Centre for Astronomy,
 Nicholaus Copernicus University,
 Gagarina 11, 87--100 Toru\'n, Poland}
\address{\P\ {\sc laco}, Facult\'e de Sciences,
123 avenue Albert Thomas, 87060 Limoges Cedex, France
}
\begin{abstract}
  We investigate a generalisation of the three dimensional
  spring-pendulum system. The problem depends on two real parameters
  $(k,a)$, where $k$ is the Young modulus of the spring and $a$
  describes the nonlinearity of elastic forces.  We show that this
  system is not integrable when $k\neq -a$. We carefully investigated
  the case $k= -a$ when the necessary condition for integrability
  given by the Morales-Ramis theory is satisfied. We discuss an
  application of the higher order variational equations for proving
  the non-integrability in this case.
\end{abstract}
\pacs{05.45.-a,02.30.Hq,45.20Jj}
\submitto{\JPA}
\eads{\mailto{maciejka@astro.ia.uz.zgora.pl},\ \mailto{Maria.Przybylska@sophia.inria.fr},\ \mailto{Jacques-Arthur.Weil@inria.fr}}
\maketitle
\section{Introduction}

The spring-pendulum, which is known also under other names: swinging
spring or elastic pendulum, is a very simple mechanical system having
a very complex dynamical behaviour and this is why sometimes it is
included to nonlinear paradigms. It consists of a point with mass $m$
suspended from a fixed point by a light spring, moving under a
constant vertical gravitation field.  In Cartesian coordinates
$(x,y,z)$ with the origin at the point of suspension of the pendulum,
the system is described by the following Hamiltonian
\begin{equation}
\label{eq:H0}
H_0=\frac{1}{2m}(p_x^2+p_y^2+p_z^2)+ mgz+\frac12k(r-l_0)^2,
\end{equation}
where $r=\sqrt{x^2+y^2+z^2},$ $l_0$ is the unstretched length of the
spring, $k\in\R^{+}$ is the Young modulus of the spring.  The
motion of this system is a complicated combination of two motions:
swinging like a pendulum and bouncing up and down like a spring.

According to our knowledge, this system appeared first in
\cite{Vitt:33::} as a simple classical analogue for the quantum
phenomenon of Fermi resonance in the infra-red spectrum of carbon
dioxide. More about the history of this system can be found in
\cite{Lynch:02::b}.  Recently it has been analyzed in  connection with
the modelling of phenomena in the atmosphere
\cite{Lynch:01::b,Lynch:02::a,Holm:02::}. Because of the 
complicated dynamics,  various approaches for its analysis were applied:
asymptotic methods \cite{Heinbockel:63::}, various perturbation
methods \cite{Nayfeh:73::,Breitenberger:81::}, numerical methods
\cite{Nunez:90::}, various formulations of KAM theorem, the Poincare
section, the Lapunov exponents \cite{Cuerno:92::}, the Melnikov method
\cite{Alvarez:93::,Banerjee:96::}, etc. A brief review of a large number of
earlier papers on the spring-pendulum can be found in
\cite{Lynch:01::b} and \cite{Davidovic:96::}.

Hamiltonian system generated by~(\ref{eq:H0}) possesses two first
integrals: Hamilton function $H_0$ and the third component of the angular
momentum
\[
p_z=x\dot y-y\dot x,
\]
and for its complete integrabilty in the Liouville sense the third
first integral is missing. Numerical computations suggest that such
additional first integral does not exist and the system is chaotic.
The first rigorous non-integrability proof for this system was
obtained by Churchill \etal \cite{Churchill:96::a} by means of the
Ziglin theory \cite{Ziglin:82::a,Ziglin:83::a}. This result can be
formulated in the following theorem.
\begin{theorem}
\label{eq:ch}
If the Hamiltonian system given by Hamiltonian function $H_0$ is
integrable   with meromorphic first integrals in the Liouville sense,
then
\begin{equation}
k= \frac{1-q^2}{q^2-9},
\label{eq:warun} 
\end{equation}
where $q$ is a rational number. 
\end{theorem}

Morales and Ramis using their theory formulated in
\cite{Morales:99::c} obtained a stronger result; they restricted the
family~(\ref{eq:warun}) of values of parameter $k$ for which the
system can be integable. Namely, they proved in~\cite{Morales:01::b2}
the following.
\begin{theorem}
\label{thm:mo}
If the Hamiltonian system given by Hamiltonian function $H_0$ is
integrable with meromorphic first integrals in the Liouville sense, 
then
\begin{equation}
\label{eq:mocon}
k= - \frac{p(p+1)}{p^2+p-2},
\end{equation}
where $p$ is an integer. 
\end{theorem}

From the above theorem it easily follows that the physical
spring-pendulum with $k\in\mathbb{R}^+$ is non-integrable except for
the case $k=0$. For $k=0$ the system is integrable because of
separation of variables in the potential.

In fact, the results presented above concern the two dimensional
spring-pendulum system obtained in the following way.  If we choose
initial conditions in such a way that value of $p_z$  equals 
zero, then the motion takes place in a vertical plane and we obtain
the two-dimensional system. But the non-integrability of the
two-dimensional spring-pendulum implies immediately the
non-integrability of the three-dimensional spring-pendulum.

The aim of this paper is to investigate the integrability of the
spring-pendulum system when the elastic potential contains also a
cubic term.  In other words, we consider a generalised spring-pendulum
system described by the following Hamiltonian
\begin{equation}
\label{eq:H}
H=\frac{1}{2m}(p_x^2+p_y^2+p_z^2)+mgz+\frac12k(r-l_0)^2-\frac13a(r-l_0)^3,
\end{equation}
where  $k\in\mathbb{R}^+$ and $a\in\R$. Our main result is the following.
\begin{theorem}
\label{thm:main}
If the Hamiltonian system given by Hamiltonian
function~\emph{\eref{eq:H}} is integrable with meromorphic first
integrals in the Liouville sense,  then  $k=-a$.
\end{theorem}
In our proof of this theorem we apply the Morales-Ramis theory and
some tools of differential algebra. Basic facts from the Morales-Ramis
theory and some results concerning special linear differential
equations are presented in Section~2. We derive variational equations
and the normal variational equations for a family of particular
solutions in Section~3.  Theorem \ref{thm:main} is proved in
Sections~4 (case $a=0$) and 5 (case $a\neq0$). In Section~4 we revise
the result of Morales formulated in Theorem~\ref{thm:mo}.  Namely, we show
that for values of $k$ given by condition~\eref{eq:mocon} the system is
non-integrable except for the case $k=0$. In this section we also show two
different kinds of arguments which give rise to non-integrability of the 
classical  spring-pendulum system  when $k\geq0$.
In Section~6, we study the exceptional case $a=-k$ and conclude that the 
Morales-Ramis method yields no obstruction to integrability, 
whereas dynamical analysis seems to indicate that the system is not 
completely integrable.

The Morales-Ramis theory was applied to study the integrability of
many Hamiltonian systems, see examples in book~\cite{Morales:99::c}
and in
papers~\cite{Mondejar:99::,Ferrer:99::a,Ferrer:99::b,Saenz:00::,Boucher:00::,Nakagawa:01::,Maciejewski:01::,Maciejewski:01::i,Maciejewski:02::b,Maciejewski:03::a}.
The differential Galois approach was used also for proving
non-integrability of non-Hamiltonian systems,
see~\cite{Maciejewski:02::a,Maciejewski:01::j,Maciejewski:02::f}.
Difficulties in application of this theory can be of a different
nature but mainly depend on dimensionality of the problem and the
number of parameters.  Although it seems that the Morales-Ramis theory
gives the strongest necessary conditions for the integrability, as far
as we know, no new integrable system was found with the help of it.
For the system investigated in this paper we have a very exceptional
situation: we found a one parameter family of Hamiltonian systems for
which the necessary conditions of integrability are satisfied, but,
nevertheless, there is evidence that this family is not integrable,
either.  Another example of such family can be found
in~\cite{Maciejewski:03::a}.
\section{Theory}
\label{sec:theory}
Below we only mention basic notions and facts concerning the
Morales-Ramis theory following \cite{Morales:99::c,Churchill:96::b}.

Let us  consider a system of differential equations
\begin{equation}
\label{eq:ds}
\Dt x = v(x), \qquad t\in\C, \quad x\in M, 
\end{equation}
defined on a complex $n$-dimensional manifold $M$.  If $\varphi(t)$ is
a non-equilibrium solution of \eref{eq:ds}, then the maximal analytic
continuation of $\varphi(t)$ defines a Riemann surface $\Gamma$ with
$t$ as a local coordinate.  Together with system \eref{eq:ds} we can
also consider variational equations (VEs) restricted to $T_{\Gamma}M$,
i.e.
\begin{equation}
\label{eq:vds}
 \dot \xi = T(v)\xi, \qquad \xi \in T_\Gamma M.
\end{equation}
We can always reduce the order of this system by one considering 
the
induced system on the normal bundle $N:=T_\Gamma
M/T\Gamma$ of  $\Gamma$  \cite{Kozlov:96::}
\begin{equation}
\label{eq:gnve}
 \dot \eta = \pi_\star(T(v)\pi^{-1}\eta), \qquad \eta\in N.
\end{equation}
Here $\pi: T_\Gamma M\rightarrow N$ is the projection.  The system of
$s=n-1$ equations obtained in this way yields the so-called normal
variational equations (NVEs).  The monodromy group $\cM$ of system
\eref{eq:gnve} is the image of the fundamental group
$\pi_1(\Gamma,t_0)$ of $\Gamma$ obtained in the process of
continuation of local solutions of \eref{eq:gnve} defined in a
neighbourhood of $t_0$ along closed paths with the base point $t_0$.
By definition it  is obvious that $\cM\subset\mathrm{GL}(s,\C)$.  A
non-constant rational function $f(z)$ of $s$ variables $z=(z_1,\ldots,
z_s)$ is called an integral (or invariant) of the monodromy group if $
f(g\cdot z)=f(z) $ for all $g\in\cM$.

In his two fundamental papers \cite{Ziglin:82::a,Ziglin:83::a}, Ziglin
showed that if system \eref{eq:ds} possesses a meromorphic first
integral, then the monodromy group $\cM$ of the normal variational
equations \eref{eq:gnve} has a rational integral (an invariant
function). This result allowed him to formulate a necessary condition
for the integrability of Hamiltonian system.
 
If system \eref{eq:ds} is Hamiltonian then necessarily $n=2m$ and there
exists function $H$ on $M$ such that $\omega(v,u)=\rmd H\cdot u$ for
an arbitrary vector field $u$ on $M$ (here $\omega$ denotes a
symplectic structure on $M$). For a given particular solution
$\varphi(t)$ we fix the energy level $E=H(\varphi(t))$. Restricting
\eref{eq:ds} to this level, we obtain a well defined system on $(n-1)$
dimensional manifold with a known particular solution $\varphi(t)$.
For this restricted system we perform the reduction of order of
variational equations.  Thus, the normal variational equations for a
Hamiltonian system with $m$ degrees of freedom have dimension
$s=2(m-1)$ and their monodromy group is a subgroup of
$\mathrm{Sp}(s,\C)$. The spectrum of an element of the monodromy
group $g\in \cM \subset\mathrm{Sp}(2(m-1),\C)$ has  the form
\[
\mathrm{spectr} (g) = ( \lambda_1, \lambda_1^{-1}, \ldots, 
\lambda_{m-1}, \lambda_{m-1}^{-1}), \quad \lambda_i \in\C, 
\]
and  $g$ is called resonant if 
\[
\prod_{l=1}^{m-1} \lambda_l^{k_l} =1\quad \mathrm{for\ \ some} \quad 
(k_1, \ldots, k_{m-1})\in {\Z}^{m-1}, \quad \sum_{i=1}^{m-1}k_i\neq0.
\]

In~\cite{Ziglin:82::a} Ziglin proved the main theorem of his theory. 
Here we formulate it as in~\cite{Kozlov:96::}. 
\begin{theorem}
  Let us assume that there exists a non-resonant element $g\in\cM$. If
  the Hamiltonian system with $m$ degrees of freedom has $m$
  meromorphic first integrals $F_1=H,\ldots,F_m$, which are
  functionally independent in a connected neigbourhood of $\Gamma$,
  then any other monodromy matrix $g'\in\cM$ transforms eigenvectors
  of $g$ to its eigenvectors.
\end{theorem} 
Recently Morales-Ruiz and Ramis generalised the Ziglin approach
replacing the monodromy group $\cM$ by the differential Galois group $\cG$ of
NVEs, see \cite{Morales:99::c,Morales:01::b1}.  For a precise definition
of the differential Galois group see
\cite{Kaplansky:76::,Morales:99::c,Put:02::}. We can consider $\cG$ as
a subgroup of $\mathrm{GL}(s,\C)$ which acts on fundamental solutions
of \eref{eq:gnve} and does not change polynomial relations among them.
In particular, this group maps one fundamental solution to other
fundamental solutions. Moreover, it can be shown that $\cM\subset \cG$
and $\cG$ is an algebraic subgroup of $\mathrm{GL}(s,\C)$. Thus, it is
a union of disjoint connected components. One of them containing the
identity is called the identity component of $\cG$ and is denoted by
$\cG^0$.

Morales-Ruiz and Ramis formulated a new criterion of the
non-in\-teg\-ra\-bi\-li\-ty for Hamiltonian systems in terms of the
properties of $\cG^0$ \cite{Morales:99::c,Morales:01::b1}.
\begin{theorem}
\label{thm:MR}
  Assume that a Hamiltonian system is meromorphically integrable in
  the Liouville sense in a neigbourhood of the analytic curve
  $\Gamma$. Then the identity component of the differential Galois
  group of NVEs associated with $\Gamma$   is Abelian.
\end{theorem}
We see that assumptions in the above theorem are stronger than in
the Ziglin theorem.  Moreover, as $\cG\supset\cM$, Theorem~\ref{thm:MR}
gives stronger necessary integrability conditions than the
Ziglin criterion.

In applications of the Morales-Ramis criterion the first step is to
find a non-equilibrium particular solution, very often it lies on an
invariant submanifold. Next, we calculate VEs and NVEs.  In the last
step we have to check if $\cG^0$ of obtained NVEs is Abelian. Very
often in applications we check only if $\cG^0$ is solvable, because if
it is not, then the system is not integrable.

For some systems the necessary conditions for the integrability formulated in
Theorem~\ref{thm:MR} are satisfied, but, nevertheless, they are
non-integrable.  In such cases, to prove the non-integrability we can
use the stronger version of the Morales-Ramis theorem based on higher
orders variational equations \cite{Morales:99::c,Morales:00::a}. The
idea of higher variational equations is following. For
system~\eref{eq:ds} with a particular solution $\varphi(t)$ we put
\[
x=\varphi(t)+\varepsilon\xi^{(1)}+\varepsilon^2\xi^{(2)}+\cdots+
\varepsilon^k\xi^{(k)}+\cdots,
\]
where $\varepsilon$ is a formal small parameter. Inserting the above
expansion into equation \eref{eq:ds} and comparing terms of the same
order with respect to $\varepsilon$ we obtain the following chain of
linear inhomogeneous equations
\begin{equation}
\label{eq:vek}
\Dt\xi^{(k)} = \mathbf{A}(t)\xi^{(k)} + f_k(\xi^{(1)}, \ldots, \xi^{(k-1)}), \qquad k=1,2, \ldots,
\end{equation} 
where 
\[
\mathbf{A}(t) = \frac{\partial v}{\partial x}(\varphi(t)),
\]
and $f_1\equiv 0$. For a given $k$ equation \eref{eq:vek} is called
$k$-th order variational equation ($\mathrm{VE}_k$). Notice that
$\mathrm{VE}_1$ coincides with \eref{eq:vds}. There is an appropriate
framework allowing to define the differential Galois group of $k$-th
order variational equation, for details
see~\cite{Morales:99::c,Morales:00::a}.  The following theorem was
announced in~\cite{Morales:00::a}.

\begin{theorem}
\label{thm:ho}
  Assume that a Hamiltonian system is meromorphically integrable in
  the Liouville sense in a neighbourhood of the analytic curve
  $\Gamma$.  Then the identity components of the differential Galois
  group of the $k$-th order variational equations $\mathrm{VE}_k$ is
  Abelian for any $k\in\N$.
\end{theorem}

There is also a possibility that the differential Galois groups of an 
arbitrary order variational equations are Abelian. Then we have  
to use  another particular solution for the non-integrability proof.
 
For further considerations we need some known facts about linear
differential equations of special forms.  At first we consider the
Riemann $P$ equation \cite{Whittaker:35::}
\begin{equation} 
\eqalign{ 
\frac{\mathrm{d}^2\eta}{\mathrm{d}z^2}&+\left(\frac{1-\alpha-\alpha'}{z}+ 
\frac{1-\gamma-\gamma'}{z-1}\right)\frac{\mathrm{d}\eta}{\mathrm{d}z}\\ 
&+ 
\left(\frac{\alpha\alpha'}{z^2}+\frac{\gamma\gamma'}{(z-1)^2}+ 
\frac{\beta\beta'-\alpha\alpha'-\gamma\gamma'}{z(z-1)}\right)\eta=0, 
} 
\label{eq:riemann}
\end{equation}
where $(\alpha,\alpha')$, $(\gamma,\gamma')$ and $(\beta,\beta')$ are the 
exponents at singular points. They satisfy the Fuchs relation 
\[ 
\alpha+\alpha'+\gamma+\gamma'+\beta+\beta'=1. 
\]
We denote the differences of exponents by 
\[
\lambda=\alpha-\alpha',\qquad\nu=\gamma-\gamma',\qquad\mu=\beta-\beta'. 
\label{eq:difer}
\] 
Necessary and sufficient conditions for solvability of the identity
component of the differential Galois group of \eref{eq:riemann} are
given by the following theorem due to Kimura \cite{Kimura:69::}, see
also \cite{Morales:99::c}.
\begin{theorem} 
  The identity component of the differential Galois group of
  equation~\eref{eq:riemann} is solvable if and only if
\begin{itemize} 
\item[A:] at least one of the four numbers $\lambda+\mu+\nu$,
  $-\lambda+\mu+\nu$, $\lambda-\mu+\nu$, $\lambda+\mu-\nu$ is an odd
  integer, or
\item[B:] the numbers $\lambda$ or $-\lambda$ and $\mu$ or $-\mu$ and
  $\nu$ or $-\nu$ belong (in an arbitrary order) to some of the
  following fifteen families
\begin{center} 
\begin{tabular}{|c|c|c|c|c|} 
\hline 
1&$1/2+l$&$1/2+m$&arbitrary complex number&\\\hline 
2&$1/2+l$&$1/3+m$&$1/3+q$&\\\hline 
3&$2/3+l$&$1/3+m$&$1/3+q$&$l+m+q$ even\\\hline 
4&$1/2+l$&$1/3+m$&$1/4+q$&\\\hline 
5&$2/3+l$&$1/4+m$&$1/4+q$&$l+m+q$ even\\\hline 
6&$1/2+l$&$1/3+m$&$1/5+q$&\\\hline 
7&$2/5+l$&$1/3+m$&$1/3+q$&$l+m+q$ even\\\hline 
8&$2/3+l$&$1/5+m$&$1/5+q$&$l+m+q$ even\\\hline 
9&$1/2+l$&$2/5+m$&$1/5+q$&$l+m+q$ even\\\hline 
10&$3/5+l$&$1/3+m$&$1/5+q$&$l+m+q$ even\\\hline 
11&$2/5+l$&$2/5+m$&$2/5+q$&$l+m+q$ even\\\hline 
12&$2/3+l$&$1/3+m$&$1/5+q$&$l+m+q$ even\\\hline 
13&$4/5+l$&$1/5+m$&$1/5+q$&$l+m+q$ even\\\hline 
14&$1/2+l$&$2/5+m$&$1/3+q$&$l+m+q$ even\\\hline 
15&$3/5+l$&$2/5+m$&$1/3+q$&$l+m+q$ even\\\hline 
\end{tabular}\\[1.5ex] 
\end{center} 
Here $l,m$ and $q$ are integers. 
\end{itemize} 
\label{kimura} 
\end{theorem} 
Next we consider the Lam\'e equation in the standard Weierstrass form
\begin{equation} 
\frac{\mathrm{d}^2\xi}{\mathrm{d}t^2}=[n(n+1)\wp(t)+B]\xi, 
\label{eq:lame} 
\end{equation} 
where $n$ and $B$ are, in general, complex parameters and $\wp(t)$ is
the elliptic Weierstrass function with invariants $g_2$, $g_3$. In
other words, $\wp(t)$ is a solution of differential equation
\begin{equation} 
\dot x^2=f(x),\qquad f(x):=4x^3-g_2x-g_3=4(x-x_1)(x-x_2)(x-x_3). 
\label{eq:wp} 
\end{equation} 
We assume that parameters $n$, $B$, $g_2$ and $g_3$ are such that
\[
\Delta=g_2^3 - 27g_3^2\neq 0, 
\]
and thus equation $f(x)=0$ has three different roots $x_1$, $x_2$ and
$x_3$. All the cases when the Lam\'e equation is solvable are listed
in the following theorem, see~\cite{Morales:99::c}.
\begin{theorem}
\label{thm:lame}
The Lam\'e equation is solvable only in the following cases
\begin{enumerate} 
\item the Lam\'e and Hermite case (see e.g. \cite{Poole:36::}) for which 
  $n\in\mathbb{Z}$ and three other parameters are arbitrary, 
\item the Brioschi-Halphen-Crowford case (see e.g. 
  \cite{Baldassarri:81::,Poole:36::}). In this case\\ 
  $n+\frac{1}{2}\in\mathbb{N}$ and $B,g_2,g_3$ satisfy an  appropriate 
  algebraic equation, 
\item the Baldassarri case \cite{Baldassarri:81::}. Then 
  $n+\frac{1}{2}\in\frac{1}{3}\mathbb{Z}\cup\frac{1}{4}\mathbb{Z}\cup\frac{1}{5}\mathbb{Z}\backslash\mathbb{Z},$ 
  and there are additional algebraic conditions on $B,g_2,g_3.$ 
\end{enumerate} 
\end{theorem}
Let $\C(z)$ denote the set of complex
rational functions of $z$  and we consider the second
order differential equation 
\[
\eta''+p(z)\eta'+q(z)\eta=0,\qquad '=\frac{\mathrm{d}}{\mathrm{d}z},\quad
p(z),q(z)\in\C(z).
\]
Putting 
\[
   \eta = y \exp\left[ -\frac{1}{2} \int_{z_0}^z p(s)\, ds \right],
\]
we obtain its reduced form

\begin{equation}
\label{eq:normal}
  y'' = r(z) y, \qquad r(z) = -q(z) + \frac{1}{2}p'(z)  + \frac{1}{4}p(z)^2.
\end{equation}
For this equation its differential Galois group $\cG$ is an algebraic
 subgroup of $\mathrm{SL}(2,\C)$. The following theorem describes all
 possible forms of $\cG$ and relates them to forms of solutions of
 \eref{eq:normal}, see \cite{Kovacic:86::,Morales:99::c}.
\begin{lemma}
\label{lem:alg}
Let $\cG$ be the differential Galois group of equation~\emph{\eref{eq:normal}}.
Then one of four cases can occur.
\begin{enumerate}
\item $\cG$ is conjugated to a subgroup of the triangular group; in
  this case equation~\emph{\eref{eq:normal}} has a solution of the
  form $y=\exp\int \omega$, where $\omega\in\C(z)$,
\item $\cG$ is conjugated with a subgroup of 
\[
D^\dag = \left\{ \left[\begin{array}{ll} c & 0\\
                                0 & c^{-1}
                      \end{array}\right]  \; \biggl| \; c\in\C^*\right\} \cup 
                      \left\{ \left[\begin{array}{ll} 0 & c\\
                                c^{-1} & 0
                      \end{array}\right]  \; \biggl| \; c\in\C^*\right\}, 
\]
in this case equation~\emph{\eref{eq:normal}} has a solution of the form
$y=\exp\int \omega$, where $\omega$ is algebraic over $\C(z)$ of
degree 2,
\item $\cG$ is primitive and finite; in this case all
  solutions of equation~\emph{\eref{eq:normal}} are algebraic, 
\item $\cG= \mathrm{SL}(2,\C)$ and equation~\emph{\eref{eq:normal}}
  has no Liouvillian solution.
\end{enumerate}
\end{lemma}
For a definition of the Liouvillian solution see e.g.
\cite{Kovacic:86::}.  An equation with a Liouvillian solution we
called integrable. When case (i) in the above lemma occurs, we say
that the equation is reducible and its solution of the form prescribed
for this case is called exponential.
\begin{remark}
\label{rem:exp}
Let us assume that equation~\emph{\eref{eq:normal}} is Fuchsian, i.e.,
$r(z)$ has poles at $z_i\in\C$, $i=1,\ldots, K$ and at
$z_{K+1}=\infty$; all of them are of order not higher than 2. Then at
each singular point $z_i$ and $z=\infty$ we have two (not necessarily
different) exponents, see e.g.  \cite{Whittaker:35::}. One can show,
see~\cite{Kovacic:86::}, that an exponential solution which exists
when case (i) in Lemma~\ref{lem:alg} occurs, has the following form
\[
y = P\prod_{i=1}^K(z-z_i)^{e_i}, 
\]
where $e_i$ is an exponent at singular point; $P$ is a polynomial and,
moreover
\[ 
\deg P = -e_\infty -\sum_{i=1}^Ke_i,
\]
where $e_\infty$ is an exponent at the infinity.
\end{remark}
\begin{remark}
\label{rem:log}
If equation~\emph{\eref{eq:normal}} has a regular singular point $z_0$ with
exponents $(e_1,e_2)$ and $e_1-e_2 \not\in\Z$, then in a neighbourhood
of $z_0$ there exist two linearly independent solutions of the form
\[
y_i = (z-z_0)^{e_i} f_i(z), \qquad i=1,2,
\]
where $f_i(z)$ are holomorphic at $z_0$. If $e_1-e_2\in\Z$, then one
local solution has the above form (for the exponent with a larger real
part). The second solution can contain a logarithmic term, for details
see~\cite{Whittaker:35::}. If the logarithmic term appears, then it can
be shown that only case (i) and case (iv) in Lemma~\ref{lem:alg} can occur,
see~\cite{Boucher:00::}.
\end{remark}

\section{Particular solution and variational equations}
\label{sec:par}
Without loss of generality, choosing appropriately units of time, mass
and length, we can put $m=g=l_0=1$.  Then the Hamiltonian of the
generalised  spring-pendulum in spherical coordinates has the following
form
\[
H=\frac12\left(p_r^2+\frac{p_{\theta}^2}{r^2}+
\frac{p_{\varphi}^2}{r^2\sin^2\theta}\right)-r\cos\theta +\frac{k}{2}
\left(r-1\right)^2-\frac{a}{3}
\left(r-1\right)^3.
\]
As we can see $\varphi$ is a cyclic coordinate and $p_{\varphi}$ is a
first integral. Manifold
\[
\cN=\{(r,,\theta,\varphi,p_r,p_{\theta},p_{\varphi})\in\C^6\,|\,
\theta=\varphi=p_{\theta}=p_{\varphi}=0\}
\]
is invariant with respect to the flow of Hamilton equations generated
by $H$.  Hamiltonian equations restricted to $\cN$ have the form
\[
\dot r=p_r,\qquad \dot p_r=1-k(r-1)+a(r-1)^2,
\]
and can be rewritten as
\[
\ddot r=1-k(r-1)+a(r-1)^2. 
\]
Thus the  phase curve located on the
energy level $H|_{\cN}=E$ is given by the equation
\begin{equation}
\label{eq:ener}
E=\frac{{\dot r}^2}{2}+\frac{k}{2}(r-1)^2-\frac{a}{3}(r-1)^3-r, 
\end{equation}  
and hence, for the generic values of $E$, it is an elliptic curve when
$a\neq0$ (for $a=0$ it is a sphere). To find its explicit time
parametrisation we put
\begin{equation}
r=\frac{6}{a}x+\frac{2a+k}{2a},
\label{trafo1}
\end{equation}
then \eref{eq:ener} transforms into equation of the form \eref{eq:wp} with
\begin{equation}
\label{eq:g2g3}
g_2=\frac{k^2-4a}{12},\qquad g_3=\frac{k^3-6ak-12a^2(E+1)}{216}.
\end{equation}
For these invariants, $x(t)$ is a non-degenerated  Weierstrass function 
provided that
\[
\Delta=\frac{(4a-k^2)^3+[k^3-6ak-12a^2(E+1)]^2}{1728}\neq 0.
\]
But $\Delta=0$ only for two exceptional values of energy corresponding
to unstable and stable equilibria (we assume here that $a\neq 0$):
\begin{equation}
\label{eq:esu}
\eqalign{
E_{\mathrm{u}}&=\frac{k^3-6a(k+2a)+(k^2-4a)^{3/2}}{12a^2},\\
E_{\mathrm{s}}&=\frac{k^3-6a(k+2a)-(k^2-4a)^{3/2}}{12a^2}.
}
\end{equation}
For $E_{\mathrm{s}}<E<E_{\mathrm{u}}$ we obtain one parameter family
$\Gamma(t,E)$ of particular solutions $(r(t),0,0,p_r(t),0,0)$
expressed in terms  of the Weierstrass function and its derivative as
\begin{equation}
r(t)=\frac{6}{a}\wp(t;g_2,g_3)+1+\frac{k}{2a},\quad
p_r(t)=\frac{6}{a}\dot{\wp}(t;g_2,g_3).
\label{eq:psw}
\end{equation}

Particular solutions are single-valued, meromorphic and double
periodic with periods $2\omega_1$ and $2\omega_2$, and they have one 
double pole at $t=0$. Thus, Riemann surfaces $\Gamma(t,E)$ are tori
with one point removed.

Using first integral $p_{\varphi}$, we can reduce the order of VE by
two.  We choose the zero level of this first integral.  Let
$\eta=(R,P_R,\Theta,P_{\Theta})$ denote variations in
$(r,p_r,\theta,p_{\theta})$. Then the reduced variational equations
restricted to the level $p_{\varphi}=0$ have the form
\[
\frac{\mathrm{d}\eta}{\mathrm{d}t}=\mathbf{L}\eta,
\]
where matrix $\mathbf{L}$ is given by
\begin{equation}
\label{eq:vmat}
\mathbf{L}=\left[\begin{array}{cccc}
0&1&0&0\\
2a(r-1)-k&0&0&0\\
0&0&0&r^{-2}\\
0&0&-r&0
\end{array}\right].
\end{equation}

The normal variational equations read 
\[
\dot{\Theta}=\frac{1}{r^2}P_{\Theta},\qquad \dot{P}_{\Theta}=-r\Theta, 
\]
and can be written as
\begin{equation}
\label{eq:nve1}
\ddot{\Theta}+2\frac{\dot{r}(t)}{r(t)}\dot{\Theta}+\frac{1}{r(t)}\Theta=0.
\end{equation}
Putting $\Phi = \Theta r$, and expressing $r$ in terms of the
Weierstrass function using~\eref{eq:psw} we transform \eref{eq:nve1} to
the form
\begin{equation}
\ddot \Phi+\frac{k^2-144\wp(t)^2}{24\wp(t)+2k+4a}\Phi=0.
\label{eq:nvet}
\end{equation}
Apart from $t=0$, equation~\eref{eq:nvet} has other  singular points 
which are solutions of the equation
\[
\wp(t)=d:=-\frac{1}{12}(k+2a).
\]
If $d\not\in\{x_1,x_2,x_3\}$, then the above equation has two roots.
If $d=x_k,$ $k=1,2,3$, then this equation has one double root. If
$a=-k$, then $t=0$ is the only singular point and in this
case~\eref{eq:nvet} is the Lam\'e equation.

\section{Non-integrability of the classical spring-pendulum}

In this section we investigate the classical spring-pendulum, i.e. we
assume that $a=0$.  In this case Hamiltonian equations restricted to
manifold $\cN$ have the form
\[
\dot r=p_r,\qquad \dot p_r=1-k(r-1),
\]
and the phase curve corresponding to energy value $E$ is a sphere
\[
E=\frac{{\dot r}^2}{2}+\frac{k}{2}(r-1)^2-r.
\]
Making transformation $t \mapsto z=r(t)$ we transform
NVE~\eref{eq:nve1} to a Fuchsian equation with rational coefficients
and four singular points $z_0=0$, $z_1=z_1(E)$, $z_2=z_2(E)$ and
$z_3=\infty$, i.e., for  generic values of $E$ the transformed NVE is
a Heun equation.  However, changing $E$ we are able to make a
confluence of two singular points and for these special choices of $E$
the transformed NVE has the form of the  Riemann $P$
equation~\eref{eq:riemann}. We have two possibilities: we can chose
$E=E_1$ such that $z_1(E_1)=z_2(E_1)$, or we can take $E=E_2$ such
that $z_1(E_2)=0$. In both cases we obtain a  Riemann $P$ equation,
however these two Riemann equations are not equivalent and thus they
give two \emph{different} necessary conditions for the integrability.
It seems that this fact was not noticed in  previous
investigations.

Let us assume that $k\neq 0$ and put $E=-(2k+1)/(2k)$.  Then the following 
 change of variable 
\[
t\mapsto z:=\frac{k}{1+k} \,r(t),
\]
 transforms~\eref{eq:nve1} to the form 
\begin{equation}
y''+\left(\frac{2}{z}+\frac{1}{z-1}\right)y'+
\left(-\frac{1}{(1+k)(z-1)^2}+\frac{1}{(1+k)z(z-1)}\right)y=0, 
\label{eq:riemann2}
\end{equation}
where $y=y(z):=\Theta(t(z))$. This  Riemann $P$ equation 
has  exponents
\[
\alpha=0,\quad \alpha'=-1,\qquad \beta=2,\quad\beta'=0,\qquad \gamma=-\gamma'=
\frac{1}{\sqrt{1+k}}.
\]
The prescribed choice of the energy corresponds to $E_1$, i.e., in the
generic Heun equation two  non-zero singular points collapse to one.  We
prove the following.
\begin{lemma}
\label{lem:rp1}
If $k\neq 0$ and 
\[  
k\neq \frac{1}{(m+2)^2}-1,
\]
where $m$ is a non-negative integer, then equation~\emph{\eref{eq:riemann2}} does
not possess a Liouvillian solution.
\end{lemma} 
\begin{proof}
  Local computation shows that equation~\eref{eq:riemann2} has
  logarithms in its formal solutions at zero and infinity whenever
  $k\neq 0$. Thus, as we know from Remark~\ref{rem:log},   if the
  equation has a Liouvillian solution, then it must be an exponential
  one, i.e. we are in case (i) of Lemma~\ref{lem:alg}. 
  As the equation is Fuchsian,  from Remark~\ref{rem:exp} it follows that 
  such exponential solution has the form 
\[
y= z^{e_{0}} \, (z-1)^{e_{1}} P(z), 
\]  
where $e_{i}$ is an  exponent at $z=i$, $i=0,1$, and $P$ is a polynomial
whose degree $m$ satisfies $m=-e_{\infty}-e_{0}-e_{1}$.  Moreover, an
expansion of an exponential solution of the form given above around a
singular point does not contain logarithms.  However, we know that
there are formal solutions at $z=0$ and $z=\infty$ with logarithms. Thus
those without logarithms corresponds to the maximal exponents, see
\cite{Whittaker:35::}.  Hence, we must put $e_{0}=\alpha=0$,
$e_{\infty}=\beta=2$, and we may take
$e_{1}=\gamma=1/\sqrt{1+k}$.  The condition on degree  of  $P$
imposes that 
\[
k=\frac{1}{(m+2)^2}-1,
\]
with $m$ a non-negative integer. As we excluded such values of $k$
this finishes the proof.
\end{proof}

Now, for all non-negative integers $m$, we have $(m+2)^{-2}-1<0$
so, as for a physical spring we have $k>0$, the above lemma shows that
equation~\eref{eq:riemann2} has no exponential solution (which
was the only possible integrable case) and, finally, the NVE is not
integrable.  This ends the proof of Theorem~\ref{thm:main} for case
$a=0$.

\begin{remark}
  Of course we can prove Lemma~\ref{lem:rp1} using 
  Theorem~\ref{kimura}.  For  equation~\emph{\eref{eq:riemann2}}
  differences of exponents are
\[
\lambda=1,\qquad \nu=\frac{2}{\sqrt{1+k}},\qquad \mu=2.
\]
In is easy to notice that case B in the Kimura theorem is impossible. Thus
equation~\emph{\eref{eq:riemann2}} is solvable (i.e. the identity
component of it differential Galois group is solvable) if and only if
the condition from case A of the Kimura Theorem is satisfied.  The
four numbers from  case A of the Kimura theorem are equal to
\[
\eqalign{
\lambda+\mu+\nu&=3+\frac{2}{\sqrt{1+k}},\qquad -\lambda+\mu+\nu=1+
\frac{2}{\sqrt{1+k}},\\
\lambda-\mu+\nu&=-1+\frac{2}{\sqrt{1+k}},\qquad
 \lambda+\mu-\nu=3-\frac{2}{\sqrt{1+k}}.
}
\]
The condition that at least one of them is an odd integer is equivalent to 
$k=(m+2)^{-2}-1$. We gave another proof of Lemma~\ref{lem:rp1} in
order to demonstrate a technique which we use in the next section.
\end{remark}

To apreciate the relevance of the physical hypothesis $k>0$, we prove
the following.
\begin{lemma}
  If $k=(m+2)^{-2}-1$ with $m$ a non-negative integer, then the
  identity component of the differential Galois group of
  equation~\emph{\eref{eq:riemann2}} is Abelian.
\end{lemma}
\begin{proof}
  Proceeding as in the proof of Lemma~\ref{lem:rp1}, we conclude that
  under our assumption equation~\eref{eq:riemann2} is solvable if and
  only if it has a solution of the form $y=P/ (z-1)^{m+2}$
  with $P$ a polynomial of degree $m$.  Following the method of
  \cite{Boucher:00::}, we make the change of variables
  $y=Y/(z-1)^{m+2}$ in~\eref{eq:riemann2}  and compute the recurrence
  relation satisfied by the coefficients of a power series solution
  $Y=\sum u_{n} z^n$ at zero.  The recurrence is:
\[
    (n-m)(n-2-m) u_{n} = (n+1)(n+2) u_{n+1}.
\]
The latter always admits a solution such that $u_{-1}=u_{m+1}=0$,
$u_{0}=1$ and $u_{m}\neq 0$, which proves that for all non-negative 
integers $m$ the NVE with $k=(m+2)^{-2}-1$ admits a solution of
the form $y=P/{(z-1)^{m+2}}$ with $P$ a polynomial of degree $m$.
Thus, the differential Galois group of equation~\eref{eq:riemann2}
conjugates to a subgroup of the triangular group (case (i) in
Lemma~\ref{lem:alg}). Moreover, as all exponents are rational, its
identity component is Abelian. 
\end{proof}

The above lemma shows that when $k<0$ (so, for the negative
Young modulus) the necessary condition of the Morales-Ramis theory is
satisfied for  infinite many cases.  As integrable systems are
extremely rare, it is worth  checking if, even for non-physical
values of $k$ excluded in Lemma~\ref{lem:rp1}, the system is integrable
or not.

To answer this question we take $E=k/2$, and make the following change
of variable 
\[
t\mapsto z:= \frac{kr(t)}{2(k+1)},
\]
in equation~\eref{eq:nve1}. Choosing the prescribed value of energy,
we perform a confluence of one non-zero singular point with $z=0$ in
the generic NVE, i.e. this energy corresponds to $E_2$.  The NVE takes
the following form
\begin{equation}
y''+\left(\frac{5}{2z}+\frac{1}{2(z-1)}\right)y'+
\left(\frac{1}{2(1+k)(z-1)^2}-\frac{1}{2(1+k)z(z-1)}\right)y=0.
\label{eq:riemann1}
\end{equation}
This is exactly the form of NVE which appears in papers
\cite{Churchill:96::a,Morales:99::c,Morales:01::b2} and the condition for
its non-integrability is given by \eref{eq:mocon}.  Combining the
non-integrability conditions for equations~\eref{eq:riemann2} and
\eref{eq:riemann1} we show the following.
\begin{theorem}
  The classical spring-pendulum system given by
  Hamiltonian~\eref{eq:H0} with $k\in\R$ is integrable only when
  $k=0$.
\end{theorem}
\begin{proof}
  Assume that the system is integrable. Then both NVEs
  \eref{eq:riemann2} and \eref{eq:riemann1} are integrable, i.e.
  they possess Liouvillian solutions. Thus, we have
\[
k=\frac{1}{(m+2)^2}-1, 
\]
for some non-negative integer $m$, and 
\[
k= - \frac{p(p+1)}{p^2+p-2},\qquad p\in\mathbb{Z}. 
\] 
But we can rewrite these conditions in the following form
\[
k = \frac{1-s}{s}, \qquad s=(m-2)^2,
\]
and 
\[
k = \frac{r}{1-r}, \qquad r = \frac{1}{2}p(p+1).
\]
As we assumed that $k\neq0$, both $s$ and $r$ are positive integers.
Now, from equality
\[
  \frac{1-s}{s}= \frac{r}{1-r},
\]
it follows that $r+s=1$, but it is impossible for positive integers $r$
and $s$.
\end{proof}

\section{Non-integrability of the generalised  spring-pendulum in the case  
$a\neq0$ and $a\neq-k$}

NVE given by~\eref{eq:nvet} depends on the energy $E$
through the invariants  of the Weierstrass function, see
formula~\eref{eq:g2g3}. The choice of the value of energy is relevant 
for computation and we put
\[
\label{eq:E}
E =E_0:= \frac{2(3k+2a)a^2-1}{12a^2}.
\]
For this value of the energy we have the following.
\begin{lemma}
    If $a\neq 0$ and $a\neq -k$, then the differential Galois group of 
    the normalized NVE \eref{eq:nvet}  for $E=E_0$ is equal to 
    $\mathrm{SL}(2,\C)$.
\end{lemma}

\begin{proof}
  Computation shows that the image of equation~\eref{eq:nvet} under
  the change of variable $t\mapsto x=\wp(t)$ is the following
\begin{equation}
\label{eq:yx}
 y''(x)
        + \frac{1}{2} \frac{f'(x)}{f(x)}y'(x)- 
 \frac {144 x^2 -k^2 - 2 a ( a+k )  }
{ \left( 12x+k+2a \right) f(x) } y(x) =0, 
\end{equation}
where 
\[
f(x)=4\,x^3-g_{2}x-g_{3} = 4(x-x_1)(x-x_2)(x-x_3),
\]
$g_2$ and $g_3$ are given by~\eref{eq:g2g3} with $E=E_0$, and $y(x) =
\Phi(t(x))$.  This equation is  Fuchsian and it has five singular points:
$x_{0}=-(2a+k)/12$, the three roots $x_1$,$x_2$, $x_3$ of $f(x)$, and
$x_4=\infty$.  The exponents at the first singularity $x_{0}$ are
$(0,1)$, the exponents at the roots of $f(x)$ are $(0,{1}/{2})$,
and the exponents at infinity are $(-1,{3}/{2})$.

If $a\not\in\{0,-k \} $, then calculation of the formal solutions at
$x_{0}$ shows that they contain a logarithm. So from
Remark~\ref{rem:log} we know that the differential Galois group of
equation~\eref{eq:yx} is either reducible or it is 
$\mathrm{SL}(2,\C)$.

Let us first assume that the equation is reducible (case (i) of
Lemma~\ref{lem:alg}), i.e. it has an exponential solution. 
From Remark~\ref{rem:exp} we know that such solution has 
the form 
\[
y=P(x) \prod_{i=0}^3(x-x_{i})^{e_{i}},
\]
where the $e_{i}$ is an exponent at $x=x_{i}$, and the degree $m$ of
polynomial $P(x)$ satisfies $m=-e_{\infty}-\sum_{i=0}^3 e_{i}$.  

Because the formal solution of valuation 0 at $x_{0}$ has a logarithm,
the valuation (i.e. the exponent) of $y$ at $x_0$ must be equal to $1$,
so $e_0=1$.  The exponents at $x_i$ for $i=1,2,3$ of $f$ are of the
form $n_{i}/2$, ($n_i$ a non-negative integer), so the relation for
the degree $m$ is either $m=-3/2-1-n/2$, or $m=1-1+n/2$, for some
non-negative integer $n= n_1+n_2 +n_3$.
   
If $m=-3/2-1-n/2$, then $m<0$, which is not possible, so we must
have $m=1-1-n/2=-n/2 $, which is possible only if
$n=0$, and hence $m=0$.
   
So the only possibility is $y=(x-x_{0})$. Substituting this candidate
into equation~\eref{eq:yx} shows that this is not a solution. Hence, the
 equation
is irreducible and, because of the logarithms in the local solutions,
the only possibility is that the differential Galois group is the full
$\mathrm{SL}(2,\C)$, which proves the lemma.
\end{proof}
  
Our main Theorem~\ref{thm:main} in the Introduction now follows, as an
immediate consequence of the Morales-Ramis theorem, from the considerations
 in Section~4 (for the case $a=0$) and from the above lemma. 

\section{Non-integrability of the generalised spring-pendulum in the 
case $a=-k$}

First we show that for the excluded case $a=-k$ the necessary
condition for  integrability given by the Morales-Ramis theory is
fulfilled.
\begin{lemma}
  For generalised spring-pendulum in case $a= -k$ the identity
  component of differential Galois group of NVE~\emph{\eref{eq:nvet}} 
  is Abelian.
\end{lemma}
\begin{proof}
For $a=-k$ equation~\eref{eq:nvet} reads
\begin{equation}
\label{eq:lamaphi}
\ddot \Phi = \left(6\wp(t) +\frac{1}{2}k\right)\Phi,
\end{equation}
so, it has the form of the Lam\'e equation \eref{eq:lame} with $n(n+1)=6$, and
$B=k/2$.  For the prescribed choice of parameters the invariants $g_2$
and $g_3$ of the Weierstrass function $\wp(t)$ are following

\[
g_2=\frac{(k+4)k}{12},\qquad g_3=\frac{k^2(k-12E-6)}{216}.
\]
The discriminant
\[
\Delta=\frac{k^3[k(k-12E-6)^2-(k+4)^3]}{1728},
\]
is only zero for two exceptional values of $E$ corresponding to two local
extrema of the potential. Assuming that $E$ is different from these
exceptional values, we can apply Theorem~\ref{thm:lame}.

As for equation~\eref{eq:lamaphi} we have $n(n+1)=6$, so $n=2$ or
$n=-3$.  Since $n\in\Z$ Lam\'e equation~\eref{eq:lamaphi} is solvable
and possesses the Lam\'e-Hermite solutions
\cite{Morales:99::c,Poole:36::,Whittaker:35::}. But for a Lam\'e
equation with such solutions the differential Galois group is Abelian
\cite{Morales:99::c}. 

For the excluded energy values, i.e. when $E=E_{\mathrm{s}}$ or
$E=E_{\mathrm{u}}$ (see formula~\eref{eq:esu}) the NVE (after
transformation $t\mapsto z:=r(t)$) has the form of Riemann $P$
equation which is solvable. Namely, for both choices of $E$ the case
(ii) from Lemma~\ref{lem:alg} occurs. Thus, the identity component of
the differential Galois group of NVE is Abelian.
\end{proof}

Let us notice here that we have at our disposal another family of
particular solutions corresponding to the following invariant manifold
\[
\cN_1=\{(r,\theta,\varphi,p_r,p_{\theta},p_{\varphi})\in\C^6\,|\,
\varphi=p_{\theta}=p_{\varphi}=0, \theta=\pi\}.
\]
However, as calculations show, using these particular solutions we
do not obtain new  necessary conditions for the integrability. 
Because of this,   we decide to apply Theorem~\ref{thm:ho}. 

 Following the decoupling of the 
first VE into tangential and normal equations (see \eref{eq:vmat}),
we find that the second variational equations are the following,
\begin{equation}
\label{nve2:1}
\ddot{r}_{2}
- 12 \, \wp\, r_2
={\frac {8{k}^{3} \left( p_{\theta,1} \right) ^{2}}
{ \left( -12\,\wp+k \right) ^{3}
}}
-k \left( r_1  \right) ^{2}
-\frac{1}{2}\, \left( \theta_{1}  \right) ^{2},
\end{equation}
and
\begin{equation}
\label{nve2:2}
\eqalign{
&\ddot{\theta}_{2}
-\,{\frac {24 \dot{\wp}\, } {-12\,\wp+k}} \dot{\theta}_2
+{\frac { 2 \,k }{-12\,\wp+k}} \theta_2
=  
\,{\frac { 16 {k}^{3} r_{1} p_{\theta,1}}{
 \left( 12\,\wp-k \right) ^{3}}}\\
&+\,
 \frac {16 {k}^{3} p_{\theta,1}\dot{r}_{1} }
          { \left( 12\,\wp-k \right) ^{3} }
          -\,
 \frac { 192 {k}^{3} \dot{\wp} \, }
           { \left( 12\,\wp-k \right) ^{4} } r_{1} p_{\theta,1}
           -
\,\frac { 4 \theta_{1} {k}^{2} r_{1} }{ \left( 12\,\wp-k \right) ^{2} },   
}
\end{equation}
where $(r_{1},\theta_{1},p_{r,1},p_{\theta,1})$ refer to solutions of
the first variational system and
$(r_{2},\theta_{2},p_{r,2},p_{\theta,2})$ refer to solutions of the
second variational system that we want to solve.  The equations are
now inhomogeneous, with left-hand sides corresponding to the
(homogeneous) first variational equations, and right-hand sides formed
of solutions of the first variational equations (which induces
coupling).

These equations look non-linear, at first. However, as explained in
\cite{Morales:99::c}, the right-hand sides are formed of linear
combinations of solutions of the second symmetric powers of the first
variational system. Hence, the second variational system, together
with the first, still reduces to a linear differential system and it
makes sense to study its differential Galois group and its
integrability. This fact remains true for variational equations of an
arbitrary order.

As the first variational equations are solvable, we could write explicit 
solutions and then solve the second variational equations by variation
of constants, but a better strategy is to proceed as in 
\cite{Morales:00::a}: as the first variational equations are Lam\'e 
equations, they have Abelian Galois group
if and only if their formal solutions at zero do not contain logarithms 
\cite{Morales:99::c}, and it is shown in \cite{Morales:00::a} that this 
remains true for variational equations of an  arbitrary order.

This is easily tested in the following way: first we compute formal
solutions (as a power series) of the first variational equations around
zero.  Then we plug a generic linear combination of these power series
in the right hand sides of \eref{nve2:1} and \eref{nve2:2}. Next we  apply
the method of variation of constants: we thus have to integrate a (known)
combination of power series and there is a logarithm if and only if
this combination of power series has a non-zero residue (i.e. a term of
degree $-1$ in its (Laurent) expansion in powers of $t$).

Performing this computation we show that the second variational
equations are integrable. Iterating the process, we computed the
solutions of the third, fourth, \ldots , until the seventh variational
equations and found that they are all integrable.  We could not
continue the calculations to higher variational equations for the
following reasons.

The first fact is that the size of the right hand sides of the
successive variational equations grows rapidly.

The second fact is that the valuation of the solutions decreases as the
order of the variational equation grows.  For example, the valuations
of $r_{2}$ is $-4$, the valuation of $r_{3}$ is $-5$,.., and the
valuation of $r_{7}$ is $-9$.  To obtain $r_{6}$ with an accuracy up
to the term of degree $0$ (to obtain the terms of negative valuation
properly, which is all we need for integrability by the above
remarks), we need to start from an $r_{1}$ with 27 terms. To obtain
$r_{7}$, we need to start from an $r_{1}$ with 30 terms, and so on.
The combination of these two facts makes the computation
intractable for  the  variational equations of order eight.

The fact that the valuations decrease is no surprise. We know that the
restriction of the Hamiltonian flow to the invariant manifold $\cal N$
of Section~\ref{sec:par} is an integrable system with one degree of
freedom.  Calculations show that the corresponding solution
$(r,0,p_{r},0)$ has a valuation at zero that decreases just like the
$r_{i}$ above (and indeed seems to govern the lowest valuation in the
$r_{i}$).
\begin{figure}[th]
\centering{ \includegraphics[scale=1]{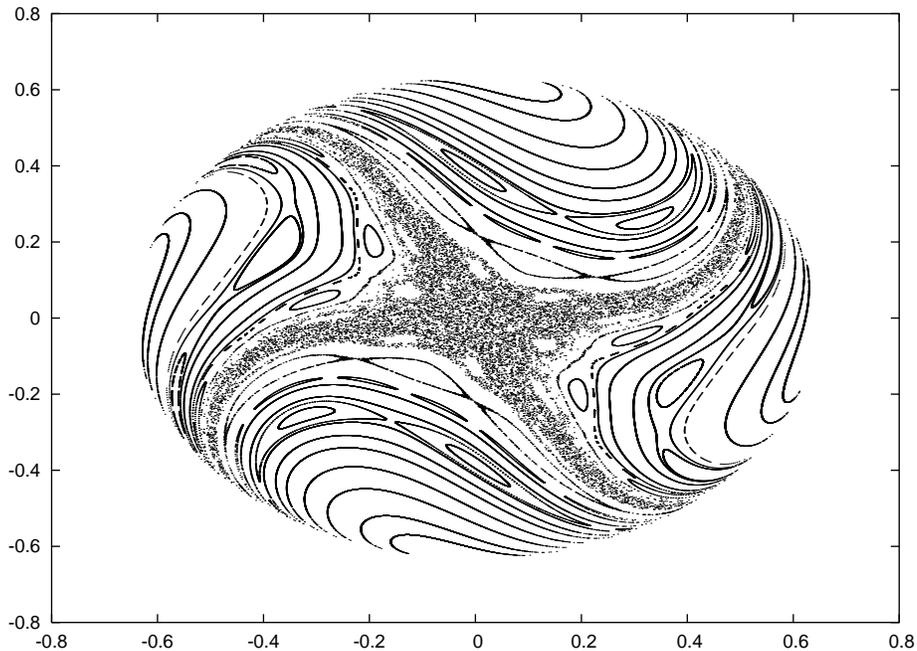}}
\caption{\label{fig:cross} 
  Poincar\'e cross section for the generalised spring-pendulum when
  $k=-a=4/3$ and $E=-0.8$. The cross section plane with coordinates $(\theta,
  p_\theta)$ is fixed at  $r=1$. }
\end{figure}

Now the fact that the variational equations up to order 7 are
integrable might lead to a suspicion that the system could be
integrable.  However, numerical experiments clearly indicate chaotic
behaviour which contradicts meromorphic integrability. We show an
example of our numerical experiments in Figure~\ref{fig:cross}. In
this figure we show the Poincar\'e cross section for energy $E=-0.8$
and $k=-a=4/3$. On the level $H=E$ we chose $(r,\theta,p_\theta)$ as
coordinates. The cross-section plane was fixed at $r=1$.

The model of the generalised swinging pendulum for $a=-k$ is thus a
puzzling example of a system that seems (numerically) to be
non-integrable but where even a deep application of the Morales-Ramis
theory is not enough to detect rigorously this non-integrability.
There is only one reported result concerning application of higher
variational equations for proving non-integrability.  In
\cite{Morales:00::a} Morales-Ruiz  reports that for a certain case
of the Henon-Heiles system the identity component of the differential
Galois group of first and second order VEs is Abelian but for the
third order VE it is not. We have also several examples of Hamiltonian
system for which $\mathrm{VE}_k$ have Abelian identity component of
differential Galois group for $k<3$ but non-Abelian for $k=3$. Thus,
as far as we know the generalised spring-pendulum system with $a=-k$
is the only example where the application of higher order variational
equations is unsuccessful in proving non-integrability. 

\ack 
We thank Martha Alvarez-Ramirez i Joaqu{\'{\i}}n Delgado for
sending us reprints of their papers.  As usual, we thank Zbroja not
only for her linguistic help. For the second author this research has
been supported by a Marie Curie Fellowship of the European Community
programme Human Potential under contract number HPMF-CT-2002-02031.

\section*{References}
\newcommand{\noopsort}[1]{}\def\cprime{$'$} \def\cprime{$'$} \def\cdprime{$''$}

\end{document}